\documentstyle[aps,epsf]{revtex}
\begin{document}
\twocolumn[\hsize\textwidth\columnwidth\hsize\csname@twocolumnfalse\endcsname
\title{Sliding Columnar Phase of DNA-Lipid Complexes}
\author{C.S. O'Hern and T.C. Lubensky}
\address{Department of Physics and Astronomy, University of Pennsylvania,
Philadelphia PA 19104}
\maketitle
\begin{abstract}
We introduce a simple model for DNA-cationic-lipid complexes in which
galleries between planar bilayer lipid lamellae contain DNA 2D smectic
lattices that couple orientationally and positionally to lattices in
neighboring galleries.  We identify a new equilibrium phase in which there
are long-range orientational but not positional correlations between DNA
lattices.  We discuss properties of this new phase such as its x-ray
structure factor $S({\bf r})$, which exhibits unusual
$\exp (- {\rm const.}\ln^2 |{\bf r}|)$ behavior as a function of in-plane
separation ${\bf r}$.
\end{abstract}
\vskip2pc]
\par
\begin{figure}
\centerline{\epsfbox{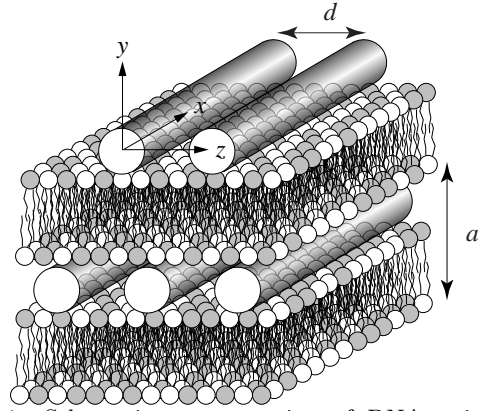}}
\caption{Schematic representation of DNA-cationic-lipid complex.  Parallel
strands of DNA form smectic lattices with lattice spacing $d$ in galleries
between lipid bilayers with spacing $a$.  Charged and neutral lipid heads
are, respectively, shaded and unshaded.  DNA strands are aligned
parallel to the $x$ axis, and the $y$ axis is the normal to the lipid
planes.}
\label{fig1}
\end{figure}

DNA is a remarkable polymer that
exhibits a
complex phase behavior as a function of packing density, salt concentration,
and other variables\cite{Livolant91}.  It is anionic, giving up positive
counter
ions to solution.  Mixtures of DNA and cationic and neutral lipids in water
form complexes that facilitate transfection of DNA into living cells and
that play an important role in the emerging field of gene
therapy\cite{Felgner87}.
Recent x-ray experiments\cite{RadKol97}
reveal the structure of
these complexes at length scales from
$10$ to several hundred Angstroms, particularly near the isoelectric point
where the total charge of counter ions given up by the DNA equals that given
up by the cationic lipids.  The lipids form bilayer membranes that stack in a
lamellar structure (Fig.\ \ref{fig1}).  Parallel strands of DNA arrange in 2D
smectic structures in the galleries between lipid bilayers.  The distance
between lipid bilayers is equal to the diameter of a DNA molecule plus a
hydration layer.  In addition, the distance $d$ between DNA strands increases
with increasing concentration of neutral lipids in a manner consistent with
counter ions being expelled to solution and charge neutrality of the complex
being determined only by the DNA and cationic lipids.
The best fit to x-ray diffraction data is obtained when some correlation
between DNA lattices in different galleries is introduced.
\par
We undertake here a theoretical investigation of possible
equilibrium phases of these lamellar DNA-lipid complexes.
We identify a new phase, with a nonvanishing smectic compression modulus
$B$ in each gallery, in which there is long-range
orientational but not positional correlation between DNA lattices in different
galleries.  This phase exhibits no restoring force
for sliding DNA lattices rigidly relative to each other, but it does exhibit a
restoring force preventing their relative rotation.  We will refer to it
as a {\it sliding columnar phase}.  It is distinct from both the
columnar phase in which the DNA segments form a 2D lattice and the totally
decoupled phase in which there is no communication between different DNA
lattices. It is similar to the decoupled phase of stacks
of tethered membranes\cite{Toner90}.
Dislocations can destroy positional correlations
within the DNA lattices, melt the sliding columnar phase,
and produce a nematic lamellar phase with $B=0$.  Whether
they always melt the phase will be discussed in
detail elsewhere\cite{OheLub98}.
\par
We consider a model in which the DNA strands are confined
to galleries between lipid bilayers in a perfect lamellar structure (with layer
spacing $a$) with no dislocations or other defects.
We assume the ground state of DNA strands in each gallery $n$ is
that favored by electrostatic interactions, i.e., a 2D smectic lattice with
layer spacing $d = 2 \pi / k_0$.
We take the lipid bilayers to be parallel to the $xz$
plane and the DNA strands to be aligned, on average, parallel to the $x$ axis
as
shown in Fig.\ \ref{fig1}.  For the moment, we assume that the lipid bilayers
are perfectly flat and do not fluctuate. In this case,
long-wavelength properties of the DNA lattice in gallery $n$
are described entirely
in terms of displacements $u^n ({\bf r})$ along the $z$ direction,
where ${\bf r} = (x,z) $ is a
position in the $xz$ plane.  The Landau-Ginzburg-Wilson Hamiltonian for the
complex is then a sum of independent elastic energies for each gallery
and terms coupling displacements and angles in neighboring galleries:
${\cal H}={\cal H}^{\rm el} + \sum_n ({\cal H}_n^{u} + {\cal H}_n^{\theta} )$
with
\begin{eqnarray}
{\cal H}^{\rm el} & = &\case{1}{2} \sum_n\int d^2 r [ B_{2}
 (u_{zz}^n)^2 + K_{2} (\partial_x^2 u^n)^2 ] \nonumber \\
{\cal H}_n^{\theta}&=& -  V_{\theta} \int d^2 r   \cos [2(\theta^n -
\theta^{n+1})] \label{rigidH}\\
{\cal H}_n^u & = & - V_u \int d^2 r \cos [k_0 (u^n - u^{n+1} )] ,\nonumber
\end{eqnarray}
where $\theta^n \approx \partial_x u^n$, and
$u_{zz}^n = \partial_z u^n - [(\partial_x u^n )^2 +
(\partial_z u^n)^2]/2$
is the nonlinear strain for gallery $n$.  $B_{2}$ and
$K_{2}$ are, respectively,
the $2D$ compression and bending moduli.
$V_u$ is of order $(\lambda_c^2/d) e^{- 2
\pi a/d}$ where $\lambda_c$ is the charge per unit length ($e$ per $1.7$ \AA)
of DNA. We do not yet have an estimate of $V_{\theta}$ whose dominant
origin is likely a membrane mediated interaction.
\par
If $V_u$ is sufficiently strong, the DNA strands form a
regular 2D lattice in the $yz$ plane,
and the whole complex will resemble an Abrikosov flux
lattice in a high-$T_c$ material with the magnetic field parallel to the
copper-oxide planes\cite{BlaFei94}.  It will have the symmetry of an
anisotropic discotic columnar liquid crystal\cite{deGennesProst93}.  As $V_u$
is
reduced (or temperature is increased), fluctuations can melt the columnar
DNA lattice without destroying the lipid-bilayer lamellar lattice.  What is
the nature of the melted phase?  To answer this question, we consider the
limit in which the potentials
$V_u$ and $V_{\theta}$ are small.  From the known statistical properties of
$2D$ smectics, we find that $V_u$ is irrelevant and $V_{\theta}$ is relevant
with respect to the totally decoupled phase.
\par
When the potentials $V_{\theta}$ and $V_u$ are zero, we have a stack
of decoupled $2D$ smectics, whose properties are by now well
known\cite{TonNel81-1,GolWan94}.
First we will ignore dislocations.
At length scales less than the nonlinear lengths
$l_x = K_{2}^{3/2}/T\sqrt{B_{2}}$ and $l_z = l_x^2
/\lambda$, where $T$ is the temperature and
$\lambda = \sqrt{K_{2}/B_{2}}$, fluctuations are
described by the linearized elastic Hamiltonian with the nonlinear strain
replaced by the linear strain $\partial_z u$.
At lengths scales longer than $l_x$ and $l_z$,
nonlinearities lead to renormalized bending and compression moduli
$K({\bf q})$ and $B({\bf q})$ that, respectively, diverge and vanish at small
wavenumber ${\bf q}$.  In both the harmonic and nonlinear regimes, the
displacement correlation function in each gallery can be expressed as
\begin{equation}
G({\bf q} ) = q_x^{-\eta} Q( q_z/q_x^{\mu} ) \sim
\cases{q_x^{-\eta}, & $q_z = 0$\cr
       q_z^{-\eta/\mu} & $q_x = 0$.\cr}
\end{equation}
In the harmonic
regime, $\eta = 4$, $\mu = 2$, and $TG^{-1} ( {\bf q} ) = B_{2} (q_z^2 +
\lambda^2 q_x^4 )$.  In the anharmonic regime, $\eta=7/2$, and $\mu = 3/2$.
The mean-square fluctuation
in the displacement diverges in both regimes with the lengths $L_x$ and $L_z$
of the
sample in the $xz$-plane:
\begin{equation}
\langle u^2 \rangle = \int {d^2 q\over (2 \pi)^2} G( {\bf q} ) =  L_x^{2
\alpha}
f_u^{(1)} ( L_z/L_x^{\mu} ) ,
\end{equation}
where $2 \alpha = \eta - 1 - \mu = 1$ in both regimes and $f_u^{(1)} (0) \sim
T/\sqrt{K_{2} B_{2}}$ and $f_u^{(1)}( w ) \sim \lambda^2 l_z^{-2
\alpha/\mu} w^{2 \alpha/\mu}$ as $w \rightarrow \infty$.
Alternatively the
displacement correlation function,
\begin{equation}
g_u ( {\bf r} ) = \langle [(u({\bf r} ) - u( 0 ) ]^2 \rangle =  |x|^{2 \alpha}
f_u^{(2)} ( |z|/|x|^{\mu} ) ,
\end{equation}
diverges with separation ${\bf r}$.  Angular fluctuations are non divergent :
%\begin{equation}
$\langle \theta ^2 \rangle = \int { d^2 q \over (2 \pi )^2 } q_x^2 G (
{\bf q} ) = \Lambda_x^{2 \alpha} f_{\theta} (
\Lambda_z /\Lambda_x^{\mu} )$,
%\end{equation}
where $\Lambda_x \sim 1/d \sim \Lambda_z \sim 1/d$ are the wavenumber
cutoffs in $q_x$ and $q_z$ and $f_{\theta}( 0 ) = \rm const.$ and
$f_{\theta}(w) \sim w^{2 \alpha/\mu}$ .
\par
Dislocations define another length scale $\xi^{\rm d}_{2D} = n_{\rm
d}^{-1/2}
\approx \exp (E^{\rm d}_{2D}/2T)$, where $n_{\rm d}$ is the density of
dislocations and
$E^{\rm d}_{2D}$ is the energy of a dislocation, which is finite for
smectics. At length scales less than $\xi^{\rm d}_{2D}$, the system is
described by the harmonic or the nonlinear elastic theory.  At
length scales longer than
$\xi^{\rm d}_{2D}$, dislocations melt the smectic lattice, leaving a $2D$
nematic with power-law angular correlations\cite{TonNel81-1}.
In what follows, we concentrate on the case in which
$\xi^{\rm d}_{2D}$ is the longest length scale in
the problem and in which there is a
regime in which dislocations can be ignored.
\par
The coupling energies ${\cal H}_n^{\theta}$ and ${\cal H}_n^u$
are irrelevant if they
tend to zero  at large $L_x$ and $L_z$ and
relevant if they diverge with $L_x$ and $L_z$.
If DNA lattices are totally decoupled,
$\langle{\cal H}_n^u\rangle = - V_u L_xL_z
\exp(- k_0^2\langle u^2({\bf r}) \rangle )$ tends
exponentially to zero for both harmonic and anharmonic $2D$ elasticities,
and is irrelevant. On the other hand $\langle \theta^2 ({\bf r}) \rangle$ is
finite for both linear and nonlinear elasticity, and the potential
$\langle{\cal H}_n^{\theta}\rangle = - V_{\theta} L_x L_z e^{-4\langle \theta^2
\rangle}$ diverges with $L_x L_z$ and is relevant.  When dislocations are
allowed, then $\langle\cos\theta_n\rangle$ dies off as a power law at large
$L_x$ and
$L_z$, and
$\langle{\cal H}_n^{\theta}\rangle$ may be relevant or irrelevant.  Thus, if we
ignore dislocations, the angular coupling is relevant, and the decoupled
Hamiltonian will flow to a new long-wavelength Hamiltonian with
angular but not positional coupling between layers.  In the continuum limit,
this Hamiltonian is
\begin{equation}
{\cal H} = \case{1}{2} \int d^3x [ B u_{zz}^2 + K (\partial_x^2 u)^2 + K_y
(\partial_y \partial_x u)^2 ] ,
\label{Hxy}
\end{equation}
where ${\bf x} = ({\bf r} , y)$, $u( {\bf x} ) = u^{y/a} ( {\bf r} )$,
$B = B_{2}/a$, $K = K_{2}/a$,
and the nonlinear strain is still that of a two-dimensional system. Note that
there are no terms proportional to $(\partial_y u)^2$ or $(\partial_y^2 u)^2$.
The only term involving $y$ derivatives is the one $K_y (\partial_y \partial_x
u)^2 = K_y (\partial_y \theta )^2$ coupling angles in different layers.
This Hamiltonian is invariant under transformations of the type $u( {\bf x} )
\rightarrow u( {\bf x} ) + f (y )$ for any function $f(y )$ independent of
${\bf r}$,
i.e., it is invariant under arbitrary rigid displacements of one layer
relative to another.
To lowest order in $V_{\theta}$,
$K_y = V_{\theta} a \langle \cos 2\theta\rangle_0^2 = V_{\theta} a
e^{-4\langle\theta^2\rangle}$.
\par
The elastic constant $K_y$ introduces a new length $l_y = a
\sqrt{K/K_y}$.  At length scales within a gallery less than $l_y$,
DNA lattices behave like independent $2D$ smectics.  If $l_y <
l_{x,z}$, there will be a crossover from $2D$ harmonic to $3D$
sliding lattice at length scales of order $l_y$.  If $l_y > l_{x,z}$,
there will be a crossover first to $2D$ nonlinear behavior and then
to $3D$ sliding behavior.
\par
Fluctuations in $u$ in the harmonic limit are now determined by
$G({\bf q}) = T [Bq_z^2 + Kq_x^4 + K_y q_x^2 q_y^2]^{-1}$.
They diverge with system size $L$:
\begin{equation}
\langle  u^2 \rangle \sim C   \ln^2 L \,\,\,{\rm and}\,\,\,
\langle (\Delta u_p)^2 \rangle = A_p C \ln  L ,
\end{equation}
where $\Delta u_p = u^{n+p} ( {\bf r} ) - u^n ({\bf r})$,
$C = T/\sqrt{BK_y}$, and $A_p$ is a number.
Thus,
$\langle{\cal H}_n^{u}\rangle \approx - V_u L^{2 - \eta_c}$, where $\eta_c
= A_1 k_0^2 C/2$ tends to zero
at large $L$ for $\eta_c >2$, and $V_u$ is irrelevant for $ T> T_u =
(4/A_1 k_0^2)\sqrt{B K_y}$.
\par
Since $\langle u^2\rangle$ diverges with $L$, the correlation
function $g_u ( {\bf x} ) = \langle [u ( {\bf x} ) - u ( 0 ) ]^2\rangle$
diverges with ${\bf x}$.  In the limit $L \rightarrow \infty$,
$g_u ( {\bf r}, 0 )$ is finite for all ${\bf r}$
and at large $|{\bf r}|$ is
\begin{equation}
g_u ( {\bf r} , 0) \sim  C \ln^2 (\Lambda^2 \lambda |{\bf r}|) .
\end{equation}
When $y$ is nonzero, $g_u ({\bf r}, y )$ diverges with $L$
for all $y$. The behavior of $g_u ( {\bf x})$ has interesting consequences for
the x-ray structure factor $S( {\bf q} ) = \int d^3 x S ( {\bf x} ) e^{-i {\bf
q} \cdot
{\bf x}}$, where $S( {\bf x} ) = \langle e^{ik_0 [ u ( {\bf x} ) - u( 0 )
]}\rangle$.
For $y=0$, the dominant contribution to $S({\bf r}) \equiv S({\bf r}, 0)= \exp[
- k_0^2 g_u( {\bf r} , 0)/2]$ is
\begin{equation}
S( {\bf r} ) =
\cases{e^{- C k_0^2(\ln^2 \Lambda^2 \lambda |{\bf r}|)/2}, & $|{\bf r}|>
l_y$;\cr
e^{- k_0^2 |x|f_u^{(2)} ( |z|/|x|^{\mu} )/2}, & $|{\bf r}|< l_y$.\cr}
\end{equation}
The contribution to $S( 0, y)$ arising from $g_u ( 0 , y )$ is
zero in the infinite volume limit when $V_u$ is irrelevant.
There are, however,
short-range positional correlations between layers arising from the
irrelevant variable $V_u$. To lowest order in a perturbation expansion
in $V_u/T$, $S(0,y=na) \sim  (V_u/2T)^n I^n$ where
$I$ has contributions
of the form $ \int d^2 r e^{- k_0^2 g_u( {\bf r} , 0 )/2}$.
Thus, there is exponential decay of $S( 0,y ) \sim e^{-|y|/\xi_y}$ with
$\xi_y = a /\ln(2T/V_u I )$.  The correlation length $\xi_y$ is finite so
long as $V_u I <2T$.  Since $g_u ( {\bf r} , 0 )$
grows more rapidly with ${\bf r}$
in the totally decoupled phase than in the sliding phase,
$I$ and $\xi_y$ are smaller
in the latter than in the former phase.  There are
further contributions to $S( {\bf r} , y )$ that become more important as $I$
grows larger.  When $\xi_y$ diverges, there will be a transition to the
true columnar crystal phase with a nonvanishing $2D$ shear modulus.
\par
The Hamiltonian of Eq.\ (\ref{Hxy}) has nonlinear parts arising from the
nonlinear strain $u_{zz}$.  These lead to logarithmic renormalizations of the
coefficients $B$, $K$, and $K_y$ similar to those in smectic liquid
crystals\cite{GrinPel81}. We find
\begin{equation}
K_y({\bf q}) \sim K^{1/2} ({\bf q}) \sim B^{-1/3} ( {\bf q} ) \sim [\ln
(\Lambda/h(
{\bf q}))]^{1/4} ,
\end{equation}
where $h( {\bf q} ) = [q_z^2 + \lambda^2 q_x^4 + \lambda_y^2 q_x^2 q_y^2]^{1/2}
$
with $\lambda^2 = K/B$ and $\lambda_y^2 =K_y/B$.
Preservation of rotational invariance in the calculation of the above
renormalized elastic constants requires considerable care.  Details of this
calculation will be presented in a separate publication\cite{ohern}.
\par
We have thus far ignored dislocations.  In two-dimensions, dislocations cause
the smectic lattice to melt at length scales larger that $\xi_d^{2}$.
Dislocations will certainly suppress order in the sliding phase.  To
obtain quantitave estimates of their effect,
we need to calculate the energy
per unit length of a dislocation predicted by the elastic energy of the
sliding phase.  We would like to obtain not only the elastic energy of a
dislocation but also some estimate of its core energy, which can
arise either from destruction of the order parameter $\psi$, describing
periodic order, over an order-parameter coherence length $\xi$ or from
deviations of the local director (specifying in this case the normal to
the DNA strands) from the direction of preferred alignment over a bend or
twist penetration depth $\lambda_B$.  In the smectic-$A$ phase, screw
dislocations have no elastic energy; their energy is totally in the core and is
dominated in type II systems, for which $\lambda_B > \xi$, by director
misalignment\cite{typeIIdisc}.
The mass-density amplitude $\psi_{\rm DNA}$ for our DNA
lattices is strictly speaking
undefined in the sliding phase in the regions between DNA galleries.
Thus, we argue that any core energy arising from dislocation lines
parallel to the layers (i.e., in the $xz$ plane) must come from director
mismatch rather than from destruction of periodic order.  To describe
director mismatch, we consider a ``gauge" version of the elastic energy
of the sliding phase in which displacements and angles are coupled in a
rotationally invariant way:
\begin{eqnarray}
{\cal H}_g & = &\case{1}{2}\int d^3 x [ B (\partial_z u)^2 + D ( \partial_x u -
\theta)^2 \nonumber \\
& &+ K_x (\partial_x \theta)^2 + K_y (\partial_y \theta)^2
+ K_z (\partial_z \theta )^2 ] .
\end{eqnarray}
This energy reduces to the sliding phase elastic energy [Eq.\ (\ref{Hxy})]
when $\theta$ is integrated out.  It introduces twist and bend
penetrations depths $\lambda_y = \sqrt{K_y/D}$ and $\lambda_z =
\sqrt{K_z/D}$ characterizing length scales over which orientations relax
in response to layer distortions.
\par
Dislocations are topological line defects.  They can be characterized by a
density ${\bf b} ( {\bf x} )$ that determines the singular part of the
displacement
variable $u( {\bf x} )$: ${\mbox{\boldmath{$\nabla$}}}
\times  {\mbox{\boldmath{$\nabla$}}} u ( {\bf x} ) = {\bf b}({\bf x})$.
Using standard
techniques\cite{ChaLub95}, we can calculate dislocation energies in terms of
the
Fourier transform ${\bf b} ({\bf q}) $ of ${\bf b} ( {\bf x} )$:
\begin{equation}
E = {1 \over 2} B \int{d^3 q\over (2 \pi)^3}
{K({\bf q})q^2 |b_y( {\bf q} )|^2 \over
B q_z^2 + K({\bf q}) q^2 [q_x^2 + (B/D) q_z^2]} ,
\label{dislo_en}
\end{equation}
where $K({\bf q}) q^2 = K_x q_x^2 + K_y q_y^2 + K_z q_z^2$.
This energy depends only on the $y$ component of ${\bf b}$, i.e., on the
component of ${\bf b}$ normal to the bilayer planes.  When there is a single
dislocation line parallel to the $y$ axis with ${\bf b} ( {\bf x} ) = d {{\bf
e}}_y
\delta ( x ) \delta ( z)$, the energy per unit length $\epsilon$ predicted
by this equation is precisely $E^{\rm d}_{2D}/a$ where $E^{\rm d}_{2D}$ is
the energy of a two-dimensional edge dislocation.  Since Eq.\
(\ref{dislo_en}) includes orientational mismatch energy and we expect no
core energy from destruction of the mass-density order parameter, we conclude
that the energy for dislocations with ${\bf b}$ in the $xz$ plane is in fact
zero.  Thus, it is possible to form dislocation loops with arbitrarily long
segments in the $xz$ plane.
(This situation is analogous to that of the ``pancake" models of flux
lattices in high-$T_c$ superconductors\cite{FeiGes90}).
These loops are equivalent to independent $2D$ dislocations
connected by loop segments with $b_y=0$ passing between DNA galleries.
Dislocations will melt the sliding columnar phase if these
$2D$ dislocations melt the smectic lattice in any given gallery.
Smectic lattices in galleries above and below a given gallery act as
external fields orienting that gallery along the common direction of the
whole sample.  Thus, each gallery is equivalent to a smectic in an
external aligning field (whose long-wavelength Hamiltonian is the in the
universality class of the $xy$ model\cite{ChaLub95})
that will Kosterlitz-Thouless melt to a nematic at a temperature $T_N$.
If $T_N > T_u$, then for $T_u < T < T_N$, there lattices have not
melted, $V_u$ is irrelevant, and the sliding columnar phase exists.  If
$T_u > T_N$, then the sliding columnar phase only exists up to length
scales of order the KT correlation length $\xi_N = d \exp ( b /|T - T_N
|^{1/2} )$, where $b$ is a nonuniversal constant.  Beyond $\xi_N$, the
phase will be a lamellar nematic.  The possibility of $T_u
< T_N$ will be addressed elsewhere\cite{OheLub98}.
\par
In our discussion to this point, we have assumed that lipid bilayers define
absolutely rigid $2D$ galleries for the DNA.  In reality, the lipid
bilayers fluctuate.  To treat
these fluctuations, we should introduce a height variable $h^n( {\bf r} )$ for
membrane $n$ and a $y$-displacement variable $u_y^n ( {\bf r} )$
for the DNA lattice in the $n$th gallery.  Clearly the variables are
coupled, and at long wavelengths, we can assume they are locked together.
Thus, we can construct a continuum elastic hamiltonian in terms of $u (
{\bf x} ) \equiv u_z( {\bf x} )$ and $u_y ( {\bf x} )$:
\begin{eqnarray}
{\cal H}_{yz} & = & \case{1}{2} \int d^3 x [ B^z u_{zz}^2 + K_{xx}^z
(\partial_x^2
u_z )^2 + K_{xy}^z(\partial_x \partial_y u_z )^2 \nonumber \\
& & + B^y u_{yy}^2 + 2 B^{yz} u_{zz} u_{yy} \\
& & + K_{xx}^y (\partial_x^2 u_y )^2 + 2 K_{xz}^y (\partial_x \partial_z u_y
)^2 + K_{zz}^y (\partial_z^2 u_y )^2 ] . \nonumber
\end{eqnarray}
The DNA lattice introduces a preferred direction in the $xz$ plane that
causes the bend elastic constant tensor $K_{ij}^y$ for $u_y$ to be
anisotropic.  The statistical properties of $u_z$, including the
correlation function $S( {\bf q})$, predicted by this Hamiltonian are
essentially
identical to those predicted by the simpler rigid-layer Hamiltonian of Eq.\
(\ref{rigidH}).   The Grinstein-Pelcovits renormalization of $B^z$, $K_{xx}^z$
and $K_{xy}^z$ are identical to those of $B$, $K$, and $K_y$.
We can use ${\cal H}_{yz}$ to calculate smectic correlations in the
lipid bilayers determined by
$S_h ( {\bf x} ) = \langle e^{ i q_0 [ u_y ( {\bf x}
) - u_y ( 0 ) ]}\rangle$, where $q_0 = 2 \pi /a $.  X-ray diffraction
measures $S_h ( {\bf q} )$, the Fourier transform of $S_h ( {\bf x} )$, as
well as $S({\bf q})$.  The
most noticeable difference between this function and that of
standard lamellar phases arises from the anisotropy in the bending
modulus. There are other differences arising from the
$u_{zz} - u_{yy}$ coupling that will be discussed in more detail in an
upcoming publication.
\par
The intrinsic chirality of DNA molecules leads to interactions that will
cause the direction of DNA lattices to rotate, like the director in a
cholesteric phase, about
an axis perpendicular to the layers in both the sliding columnar and
lamellar nematic phases.  We expect, however, these interactions to
be small and resultant pitches to be very large
because the lipid bilayer prevents close approach of DNA
molecules.
\par
In this letter, we have introduced and investigated some of the
properties of a new phase of matter that may exist in highly anisotropic,
rotationally invariant columnar systems such as DNA-cationic-lipid
complexes.  It would be
interesting to find evidence for this phase in the DNA complexes
studied in Ref. \onlinecite{RadKol97}.
The material in these experiments
consisted of micron-size spherulites composed of apparently randomly
oriented domains of length of order $L= 500$\AA\ on a side of
DNA-lipid-complex.  The x-ray data are consistent with nearly
independent harmonic smectic layers with some exponentially attenuated
positional correlation between layers.  A nonvanishing $V_u$
will lead to exponential correlations between layers even in the harmonic
regime.  $500$\AA\ corresponds approximately to the nonlinear $2D$
length $l_x$.  Thus, it is not clear whether the domain size
is set by actual distortions of the lipid layers or by the breakdown
of harmonic behavior at length scales longer than $l_x$ or $l_z$.  It
will be necessary to prepare monodomain samples to determine whether
there is a crossover to anharmonic $2D$ behavior or to that of the
$3D$ sliding phase proposed in this paper.
\par
We are grateful to Cyrus Safinya for inspiring our
interest in this problem, to John Toner for emphasizing the importance
of dislocation interactions in the sliding phase,
and to Robijn Bruinsma and Randy Kamien for helpful discussions.
This work was supported
in part by the NSF under grant No. DMR94-23114.
\par
{\it Note added} Results essentially identical to those reported here were
obtained
independently by Golubovi\'{c} and Golubovi\'{c}
as reported in Ref. \cite{Golub}.
%\bibliography{/u/tom/datab/physics,/u/tom/datab/liqcryst}
\bibliographystyle{prsty}

\end{document}